\documentclass[aps,prl,twocolumn,superscriptaddress]{revtex4-1}

\usepackage{graphicx}
\usepackage{hyperref}

\hypersetup{
	colorlinks=true,
	linkcolor=blue,
	citecolor=blue,
	urlcolor=blue
}

\newcounter{firstbib}

\begin{document}
	
\title{Motional Quantum Ground State of a Levitated Nanoparticle from Room Temperature}	
	
\author{Uro\v s Deli\' c}
\email{uros.delic@univie.ac.at}
\affiliation{Vienna Center for Quantum Science and Technology (VCQ), Faculty of Physics, University of Vienna, Boltzmanngasse 5, A-1090 Vienna, Austria}
\affiliation{Institute for Quantum Optics and Quantum Information (IQOQI) Vienna, Austrian Academy of Sciences, Boltzmanngasse 3, A-1090 Vienna, Austria}
\author{Manuel Reisenbauer}
\affiliation{Vienna Center for Quantum Science and Technology (VCQ), Faculty of Physics, University of Vienna, Boltzmanngasse 5, A-1090 Vienna, Austria}
\author{Kahan Dare}
\affiliation{Vienna Center for Quantum Science and Technology (VCQ), Faculty of Physics, University of Vienna, Boltzmanngasse 5, A-1090 Vienna, Austria}
\affiliation{Institute for Quantum Optics and Quantum Information (IQOQI) Vienna, Austrian Academy of Sciences, Boltzmanngasse 3, A-1090 Vienna, Austria}
\author{David Grass}
\altaffiliation{Present address: Department of Chemistry, Duke University, Durham, North Carolina 27708, United States}
\affiliation{Vienna Center for Quantum Science and Technology (VCQ), Faculty of Physics, University of Vienna, Boltzmanngasse 5, A-1090 Vienna, Austria}
\author{Vladan Vuleti\' c}
\affiliation{Department of Physics and Research Laboratory of Electronics, Massachusetts Institute of Technology, Cambridge, Massachusetts 02139, USA}
\author{Nikolai Kiesel}
\affiliation{Vienna Center for Quantum Science and Technology (VCQ), Faculty of Physics, University of Vienna, Boltzmanngasse 5, A-1090 Vienna, Austria}
\author{Markus Aspelmeyer}
\email{markus.aspelmeyer@univie.ac.at}
\affiliation{Vienna Center for Quantum Science and Technology (VCQ), Faculty of Physics, University of Vienna, Boltzmanngasse 5, A-1090 Vienna, Austria}
\affiliation{Institute for Quantum Optics and Quantum Information (IQOQI) Vienna, Austrian Academy of Sciences, Boltzmanngasse 3, A-1090 Vienna, Austria}

\begin{abstract}
	We report quantum ground state cooling of a levitated nanoparticle in a room temperature environment. Using coherent scattering into an optical cavity we cool the center of mass motion of a $143$ nm diameter silica particle by more than $7$ orders of magnitude to $n_x=0.43\pm0.03$ phonons along the cavity axis, corresponding to a temperature of $12~\mu$K. We infer a heating rate of $\Gamma_x/2\pi = 21\pm 3$ kHz, which results in a coherence time of $7.6~\mu$s -- or $15$ coherent oscillations -- while the particle is optically trapped at a pressure of $10^{-6}$ mbar. The inferred optomechanical coupling rate of $g_x/2\pi = 71$ kHz places the system well into the regime of strong cooperativity ($C \approx 5$). We expect that a combination of ultra-high vacuum with free-fall dynamics will allow to further expand the spatio-temporal coherence of such nanoparticles by several orders of magnitude, thereby opening up new opportunities for macrosopic quantum experiments.  
\end{abstract}

%\keywords{}

\maketitle

The possible quantum behavior of macroscopically sized objects has puzzled researchers since the early days of quantum theory \cite{RN1}. It is still an open question whether quantum laws are universally valid or whether a classical, “macrorealistic” description of nature has to take over for sufficiently large systems \cite{RN2}. One goal of macroscopic quantum experiments is to shed light on this question by observing and controlling quantum behaviour at macroscopic scales \cite{RN3,RN4}. This task is becoming increasingly difficult as larger systems are harder to shield from environmental influences that may decohere the desired quantum phenomena \cite{RN5}. It has been suggested that levitation in high vacuum can provide a radically new platform for exploring macroscopic quantum phenomena \cite{RN6,RN7,RN8}. Levitation allows for sufficient isolation of the center-of-mass motion of solid-state objects, which enables a combination of quantum control with free-fall dynamics even at room temperature. Here we report a first relevant step by cooling an optically levitated solid, a silica sphere of 143nm diameter, by more than 7 orders of magnitude to its quantum ground state of motion in a room-temperature environment. We use cavity cooling by coherent scattering \cite{RN9,RN10}, a technique originally developed in the context of atomic physics for laser cooling of atoms and molecules \cite{RN11,RN12}, to cool the particle motion along the cavity axis to $n_x=0.43\pm0.03$  phonons, corresponding to a temperature of $12~\mu$K. We estimate ground state coherence times of approximately $7.6~\mu$s at the achieved background pressure of $10^{-6}$ mbar. This corresponds to 15 coherent oscillations within the optical trap, which can be further improved by going to ultra-high vacuum. The demonstration of a pure quantum state of motion of a levitated solid-state object with $10^8$ atoms ($2\times 10^9$ a.m.u.) in a room temperature environment paves the way for generating and controlling manifestly non-classical states of complex and massive objects at unprecedented scales. This creates new opportunities for sensing applications and tests of fundamental physics.  

Optical levitation of dielectric particles works by using forces induced by laser light that are strong enough to overcome gravity \cite{RN13}. At its most fundamental level, an incoming laser polarizes the dielectric material, which in turn interacts with the radiation field of the laser. As a consequence, a particle in a tightly focused laser beam experiences a gradient force towards the intensity maximum of the beam, resulting in a three-dimensional confinement of the particle \cite{RN14}. Such "optical tweezers" have become a powerful tool to manipulate dielectric objects in isolation from other environments. For example, the ability to trap individual mesoscopic objects and even living specimen in liquid has made a lasting impact on the field of biology and biophysics \cite{RN13,RN15}. On the other hand, trapping of micron- and sub-micron sized particles in vacuum has allowed to address previously unexplored regimes for example in stochastic thermodynamics \cite{RN16}, force sensing \cite{RN17,RN18}, metrology \cite{RN19,RN20} and laboratory scale cosmology experiments such as searches for dark matter and dark energy \cite{RN21}. In the domain of quantum physics, optical trapping and cooling of atoms \cite{RN22} has enabled the study of individual atoms and quantum gases. It is also a fundamental technique for confining particles to optical lattice geometries for the study of many-body quantum phenomena \cite{RN23,RN24}.

Based on these previous achievements it is natural to consider if a combination of optical levitation of relatively large solids in high vacuum together with the methods of quantum optics will allow for a completely new regime of macroscopic quantum physics. The proposals that have been put forward suggest exactly that \cite{RN6,RN7,RN8,RN25}. Laser cooling techniques should enable to prepare a levitated solid-state particle in its quantum ground state of motion \cite{RN6,RN7,RN25}. The particle wavepacket can then be expanded and modified by a sequence of free fall, coherent manipulations and quantum measurement operations \cite{RN8,RN26,RN27}. This results in previously unattainable macroscopic superposition states of massive objects.

\begin{figure*}
	\includegraphics[width=\linewidth]{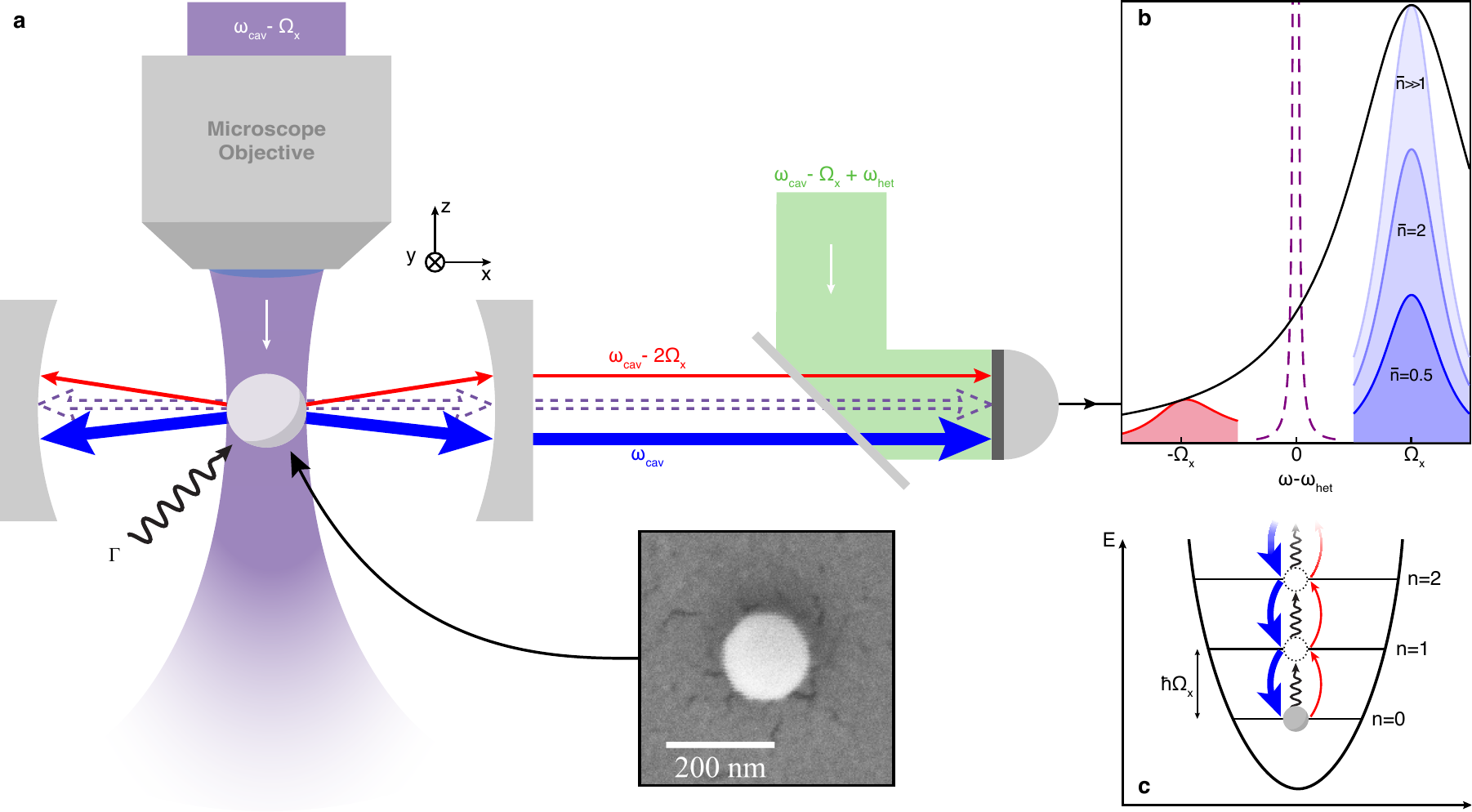}
	\caption{\label{fig:setup} \textbf{Cavity cooling and sideband thermometry.} \textbf{a,} Schematics of the apparatus for particle cooling and detection. A silica nanoparticle, shown here in the SEM image, of nominal diameter $d=143\pm4$ nm, is trapped in an optical tweezer (purple) (wavelength $\lambda=1064$ nm, waist size $W_x\approx 0.67~\mu$m and $W_y\approx0.77~\mu$m, power in the focus $P_{tw}\approx400$ mW). The frequency of the tweezer laser $\omega_{tw}$ is detuned from the initially empty cavity resonance $\omega_{cav}$ such that $\omega_{tw}=\omega_{cav}-\Omega_x$ ($\Omega_x/2\pi\approx305$ kHz is the axial motional frequency of the particle). Scattering of the tweezer light into the optical cavity is maximized via the tweezer polarization. The spatial mode overlap between the dipole emission pattern of the silica nanoparticle and the optical cavity mode results in a Purcell enhancement of the scattered radiation by a factor of approximately 8 compared to a free space configuration. When the particle is positioned at a cavity node, the elastic scattering of the tweezer light (dashed purple) is suppressed leaving only the inelastically scattered Stokes (red) and anti-Stokes (blue) sidebands at frequencies $\omega_{cav}-2\Omega_x$ and $\omega_{cav}$, respectively. Mixing the scattered sidebands with a strong local oscillator (green) at $\omega_{LO} = \omega_{tw} + \omega_\mathrm{het}$, where $\omega_\mathrm{het}/2\pi=10.2$ MHz, allows us to separately detect both sidebands in a heterodyne measurement at the cavity output port. The total heating rate from the environment is represented here as $\Gamma$. \textbf{b}, An illustration of a heterodyne measurement of the Stokes and anti-Stokes sidebands. The phonon occupation impacts the overall scattering rates which are initially modified by the cavity response (black). For large phonon occupations ($\bar{n}\gg 1$), the relative amplitudes of the Stokes and the anti-Stokes sidebands are completely described by the cavity transmission function. As the nanoparticle approaches the motional ground state, the sideband ratio is modified as the oscillator cannot undergo an anti-Stokes scattering process. This asymmetry allows for direct thermometry of the phonon occupation. The suppressed elastic scattering contribution is depicted for reference. \textbf{c}, Depiction of the phonon energy levels close to the ground state. The heating rate from the environment and the Stokes scattering rate are balanced by the anti-Stokes scattering rate.  
	}
\end{figure*} 

A key requirement for entering this new regime is to prepare the particle wavepacket in a sufficiently pure quantum state, in this case, to cool its motion into the quantum ground state. One possibility is to monitor the particle motion with a sensitivity at or below the ground state size of the wavepacket and apply a feedback force to directly counteract the motion \cite{RN28,RN29}. Such feedback cooling to the quantum ground state has recently been demonstrated for harmonic modes of cryogenically cooled micromechanical membranes \cite{RN30}. In the context of levitated nanoparticles, feedback cooling has initially been introduced to provide stable levitation in high vacuum \cite{RN31,RN32,RN33}. At present, feedback cooling is limited to approximately 4 motional quanta (phonons) \cite{RN34}. A different approach is derived from laser cooling of atoms, where the absorption and re-emission of Doppler-shifted laser photons provides a velocity dependent scattering force \cite{RN22}. The presence of an optical cavity modifies the electromagnetic boundary conditions for the scattered light, which can be used to tailor the scattering rates and allows for cooling of particles without an accessible internal level structure such as molecules or dielectric solids \cite{RN11,RN12,RN35,RN36}. These cavity-cooling schemes have been used in the past to achieve ground state cooling of various systems ranging from individual atoms to cryogenically cooled modes of solid-state nano- and micromechanical oscillators in the context of cavity optomechanics \cite{RN37,RN38,RN39}. Previous attempts to apply cavity cooling to levitated solids have proven challenging and cooling was limited to several hundred phonons \cite{RN40,RN41,RN42,RN43,RN44}, mainly due to co-trapping associated with high intracavity photon number and the excessive laser phase noise heating at the low motional frequencies ($< 1$ MHz) \cite{RN43,RN44}. We are applying a modified scheme -- cavity cooling by coherent scattering \cite{RN9,RN10,RN35,RN45} -- which circumvents these shortcomings and enables direct ground-state cooling of a solid in a room-temperature environment.

We trap a spherical silica particle with a nominal diameter of $d=143 \pm  4$ nm inside a vacuum chamber using an optical tweezer. A tightly focused laser beam (power in the focus: $P_{tw} \approx  400$ mW; wavelength $ \lambda =1064$ nm; frequency $ \omega_{tw}=2\pi c/ \lambda $ ; $c$: vacuum speed of light) creates a three-dimensional harmonic potential for the particle motion with motional frequencies ($ \Omega_{x}, \Omega_{y},\Omega_{z})/2\pi  \approx  (305,275,80)$ kHz. We position the particle within an optical cavity (cavity finesse $\mathcal{F} \approx 73.000$; cavity linewidth $ \kappa  /2 \pi =193 \pm  4$ kHz; cavity frequency $ \omega_{cav}=\omega_{tw}+\Delta$; $\Delta$: laser detuning), which collects the tweezer light scattered off the nanoparticle under approximately a right angle (FIG. 1a). The particle has sub-wavelength dimension and hence resembles to good approximation a dipole emitter -- driven by the optical tweezer the particle coherently scatters dipole radiation predominantly orthogonal to the tweezer polarization axis. Motorized translation stages in the tweezer optics allow us to position the particle with an accuracy of a few nm with respect to the cavity axis (x-direction), such that the particle can be well localized within one period of the cavity standing wave field. To achieve optimal cooling along the cavity axis the particle needs to be located at an intensity minimum of the cavity standing wave field \cite{RN9,RN10}. At that location the particle is "dark"  and accordingly all dipole scattering into the cavity mode is inhibited due to destructive interference imposed by the cavity. The particle motion breaks this symmetry and therefore only inelastically scattered Stokes- and anti-Stokes photons at sideband frequencies $\omega_{tw}\pm\Omega_x$ can propagate in the cavity.

Cavity cooling of the particle motion occurs because (Stokes-)scattering processes along the cavity, which increase the kinetic energy of the particle by $\hbar\Omega_{x}$ per photon, are suppressed, while (anti-Stokes-) scattering processes, which reduce the energy accordingly, are enhanced \cite{RN39,RN46}. This process is maximized at the optimal detuning $ \Delta\approx\Omega_{x}$, where the anti-Stokes sideband becomes fully resonant with the cavity. A particle in its quantum ground state of motion cannot further reduce its energy, hence anti-Stokes scattering close to the ground state is fundamentally inhibited (FIG.1b,c). The resulting sideband asymmetry in the scattering rates is a direct measure of the temperature of the harmonic particle motion, which does not require calibration to a reference bath\cite{RN47}. We observe these sidebands, which are modulated by the cavity envelope\cite{RN48}, using frequency-selective heterodyne detection of the cavity output, specifically by mixing it with a strong local oscillator field ($P_{LO} \approx  400~\mu$W) detuned from the tweezer laser by $ \omega_\mathrm{het}/2\pi=10.2$ MHz (FIG. 2a). Independent measurements of the cavity linewidth $\kappa$  and the laser detuning $\Delta$ allow us to correct the detected sideband ratios for the cavity envelope (see Appendix), and hence to extract the motional temperature of the particle via the fundamental sideband asymmetry (FIG. 2b). For this method to work reliably, it is important to exclude all relevant influences of noise contributions to the sideband asymmetry \cite{RN49}. We ensure this by confirming that the detection process is shot-noise limited and that both amplitude- and phase-noise contributions of the drive laser are negligible (see Appendix). Figure 2c shows the measured phonon number $n_{x}$ along the cavity axis for different laser detunings $\Delta$. For near optimal detuning of $ \Delta/2\pi=315$ kHz, we observe a final occupation as low as $n_{x}=0.43\pm 0.03$, corresponding to a temperature of $12.2\pm 0.5~\mu$K and a ground-state probability of $70\pm2\%$. Note that in contrast to previous quantum experiments involving cryogenically cooled solid-state oscillators, ground state cooling here is achieved in a room temperature environment.

\begin{figure*}
	\includegraphics[width=\linewidth]{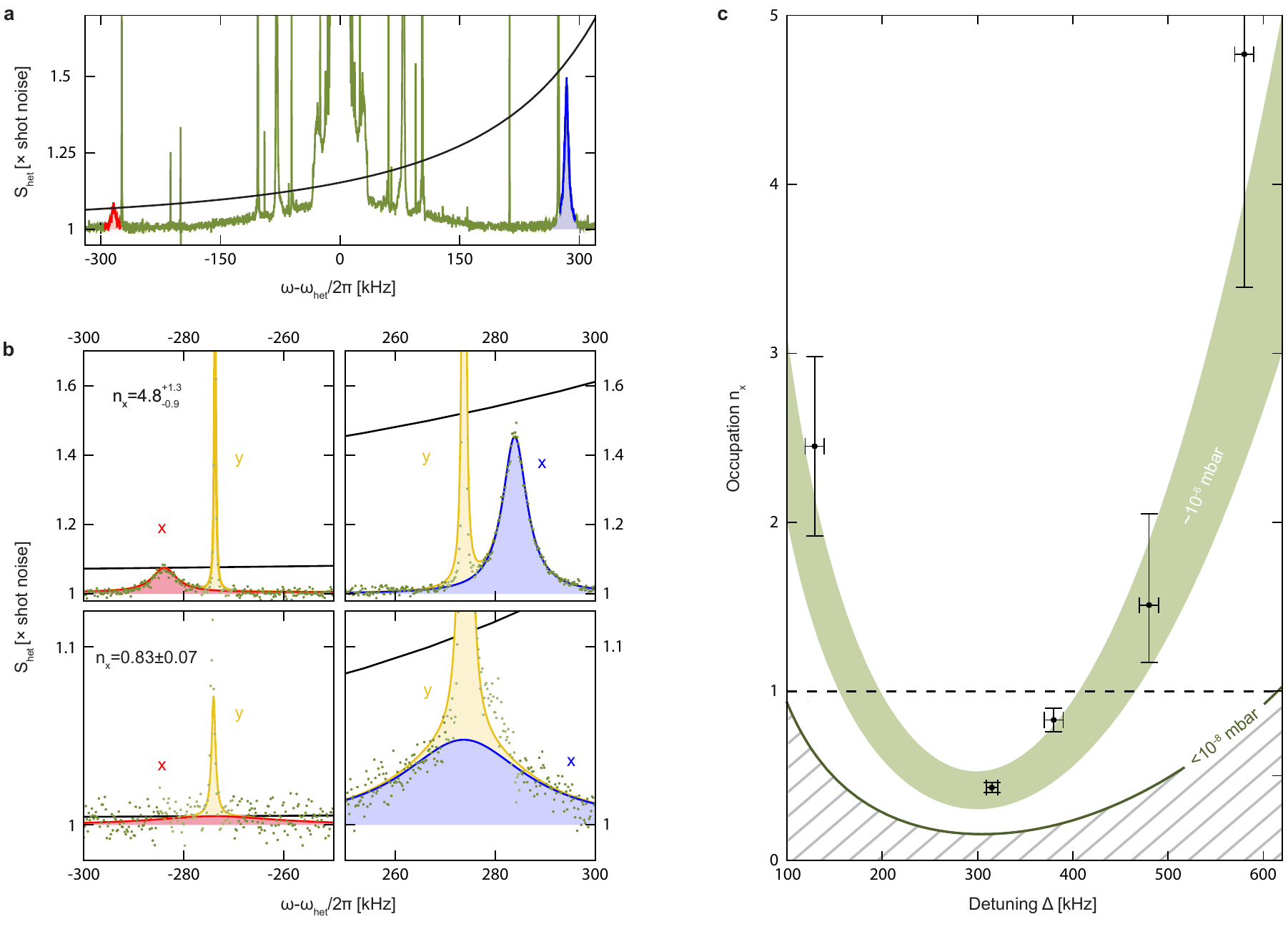}
	\caption{\label{fig:detuningscan} \textbf{Scan of the optical tweezer frequency relative to the optical cavity resonance.} \textbf{a}, Heterodyne spectrum of the Stokes and anti-Stokes sidebands. The inelastically scattered Stokes (red) and anti-Stokes (blue) sidebands are mixed with a strong local oscillator ($P_{LO}\approx 400~\mu$W) which is detuned from the optical tweezer by $\omega_\mathrm{het}/2\pi=10.2$ MHz. The resulting spectrum contains information about the motion of the nanoparticle in all 3 orthogonal directions. Highlighted in red and blue, respectively, are the features corresponding to the Stokes and anti-Stokes sidebands of the axial motional frequency $\Omega_x$. \textbf{b}, Thermometry of the phonon occupation associated with the cavity axial motion. Shown are the sideband power spectra for Stokes (left) and anti-Stokes (right) scattering for different detunings $\Delta/2\pi= 580\pm10~$ kHz (top row) and $\Delta/2\pi= 380\pm10~$ kHz (bottom row). Red and blue solid lines are fits to the data representing the $x$ motion. Also shown are yellow solid lines, which are fits to the $y$ motion at $\Omega_y/2\pi=275\pm1$ kHz. Black solid line indicates the cavity transmission function normalized to the Stokes sideband power. Smaller temperatures show a stronger deviation of the anti-Stokes scattered sideband power from the cavity envelope, as is described in detail in the main text. The ratio of amplitudes together with the independently measured cavity transmission function yields the final occupation $n_x$. \textbf{c}, Occupation nx as a function of tweezer laser detuning. The cooling rate is maximal when the optical tweezer is detuned from the cavity resonance by approximately the motional frequency $\Delta/2\pi\approx 315$ kHz. At this maximal cooling point, we achieve a phonon occupation of $n_x=0.43\pm 0.03$. The wide green band is a theoretical model based on system parameters, which takes into account pressure drifts during the measurement. The lower dark green line corresponds to the expected occupation when the environment pressure is below $10^{-8}$ mbar at which heating due to collisions with the background gas become negligible in comparison with the recoil heating.
		}
\end{figure*} 

The final occupation $n_{fin}$ along any direction is reached when the total heating rate $\Gamma_{x}$ is balanced by the cooling rate $n_{fin}\times\gamma$, where $\gamma$ is the linewidth of the motional sidebands \cite{RN50}. For the resolved sideband regime ($\kappa<\Omega_{x}$) as studied here, and in the absence of any other heating mechanisms, Stokes scattering due to the finite cavity linewidth limits cooling with optimal parameters to a minimum phonon occupation of $n_{min}=(\kappa /4\Omega_{x})^2 \approx 0.025$. Note that in this case, detailed balance implies that the fundamental ground state asymmetry exactly compensates the effect of the cavity envelope and therefore both sidebands have equal power. Additional sources of heating are balanced by larger anti-Stokes scattering, which results in the overall observed sideband imbalance. By independently measuring nx and γ we extract a total heating rate as low as $ \Gamma_{x}/2 \pi =(20.6\pm2.3)$ kHz at a pressure of $\sim10^{-6}$ mbar. This is consistent with the separately measured heating rate due to background gas collisions \cite{RN44}, $\Gamma_{gas}/2\pi =(16.1\pm1.2)$ kHz and the expected heating contributions from photon recoil,  $\Gamma_{rec}/2\pi\approx6$kHz, and laser phase noise, $\Gamma_{phase}/2\pi<200$ Hz (see Appendix). 

Using the measured heating rates we estimate a maximum coherence time of $(7.6\pm1)~\mu$s in the optical trap, corresponding to approximately 15 coherent oscillations before populating the ground state with one phonon \cite{RN6,RN27}. In a free fall experiment, where the particle would be released from the optical trap, the dominant source for decoherence is the collision with a single background-gas molecule. Note that we are operating in an interesting transition regime, in which the initial ground-state wavepacket size is smaller than the thermal de Broglie wavelength of a gas molecule by a factor of $~6$ (see Appendix). In this so-called short-distance regime the decoherence rate increases quadratically with wavepacket size \cite{RN51}. As a consequence, a freely expanding wavepacket will experience smaller coherence times than expected from the measured ground state heating rates. At the achieved background pressure of $10^{-6}$ mbar, this limits the free-fall coherence time to $~1.4~\mu$s, which would allow for an expansion of the wavepacket by approximately a factor of 3, from the ground state size of $3.1$ pm to $10.2$ pm. For wavepacket sizes much larger than the de Broglie wavelength of a gas molecule the decoherence rate saturates in the so-called long-distance regime. Significantly larger wavepacket sizes can be achieved by decreasing this decoherence rate further, for example by operating at much lower pressures. Blackbody radiation will then become the dominant source of decoherence and will, for our room temperature parameters, allow for wavepacket expansions up to several nanometres. A wavepacket size on the order of the particle radius could be achieved by combining ultra-high vacuum (approximately $10^{-11}$ mbar) with cryogenic temperatures (below 130K) (see Appendix).

Our platform also entails the ability to manipulate the spatial profile of the trapping laser for implementing nonlinear potentials, and may open the possibility to create non-classical states of motion such as non-Gaussian states or large spatial superpositions. This is in stark contrast to current experiments that prepare motional quantum states using solid state harmonic oscillators, where coupling to external non-linear systems or measurements provide the interaction for non-classical state preparation \cite{RN52,RN53,RN54}.

In conclusion, we have demonstrated ground-state cooling of a levitated solid-state object in a room temperature environment. The coherence time is limited by the background pressure of $10^{-6}$ mbar to $~7.6\mu$s, which constitutes a crucial step for the quantum state control of such massive solid-state systems. The inferred optomechanical coupling rate of $g_{x}/2\pi\approx 71$ kHz places the system well into the regime of strong cooperativity $C > 1$, with $C=4g_{x}^2/(\kappa\Gamma_{x})\approx 5$, and should already allow to implement quantum optical protocols for the generation of non-classical states of motion \cite{RN27,RN55}. In future experiments, significant reduction of decoherence can be achieved mainly by lower background pressures, but potentially also by operating at lower temperatures and using smaller cavity mode volumes. The combination of cavity optomechanical quantum control of levitated systems and free fall experiments can open up a new class of macro-quantum physics, with potential applications also to quantum sensing \cite{RN56,RN57} and other fields of fundamental physics \cite{RN58}. Most importantly, we believe that the quantum control of levitated systems is a viable route towards experiments in which quantum systems can act as gravitational source masses, as has been originally suggested by Feynman \cite{RN59} and recently been re-visited in the context of levitation \cite{RN60,RN61}.

\begin{acknowledgments}
\textit{Acknowledgments.}  We thank Oriol Romero-Isart, Lukas Novotny and Tania Monteiro for insightful comments. U.D. and M.A. thank Jun Ye for initially pointing out the relevance of coherent scattering to us. This project was supported by the European Research Council (ERC CoG QLev4G), by the ERA-NET programme QuantERA, QuaSeRT (Project No. 11299191; via the EC, the Austrian ministries BMDW and BMBWF and research promotion agency FFG), by the Austrian Science Fund (FWF, Project TheLO, AY0095221, START) and the doctoral school CoQuS (Project W1210), and the research platform TURIS at the University of Vienna.
\end{acknowledgments}

\bibliographystyle{apsrev4-1}

\setcounter{figure}{0}
\setcounter{equation}{0}
\renewcommand{\thefigure}{S\arabic{figure}}
\renewcommand{\theequation}{S\arabic{equation}} 

\newpage

\begin{center}
\LARGE{\textbf{Appendix}}
\end{center}	
\normalsize

\section{The experimental setup}

\begin{figure}[h!]
	\includegraphics[width=\columnwidth]{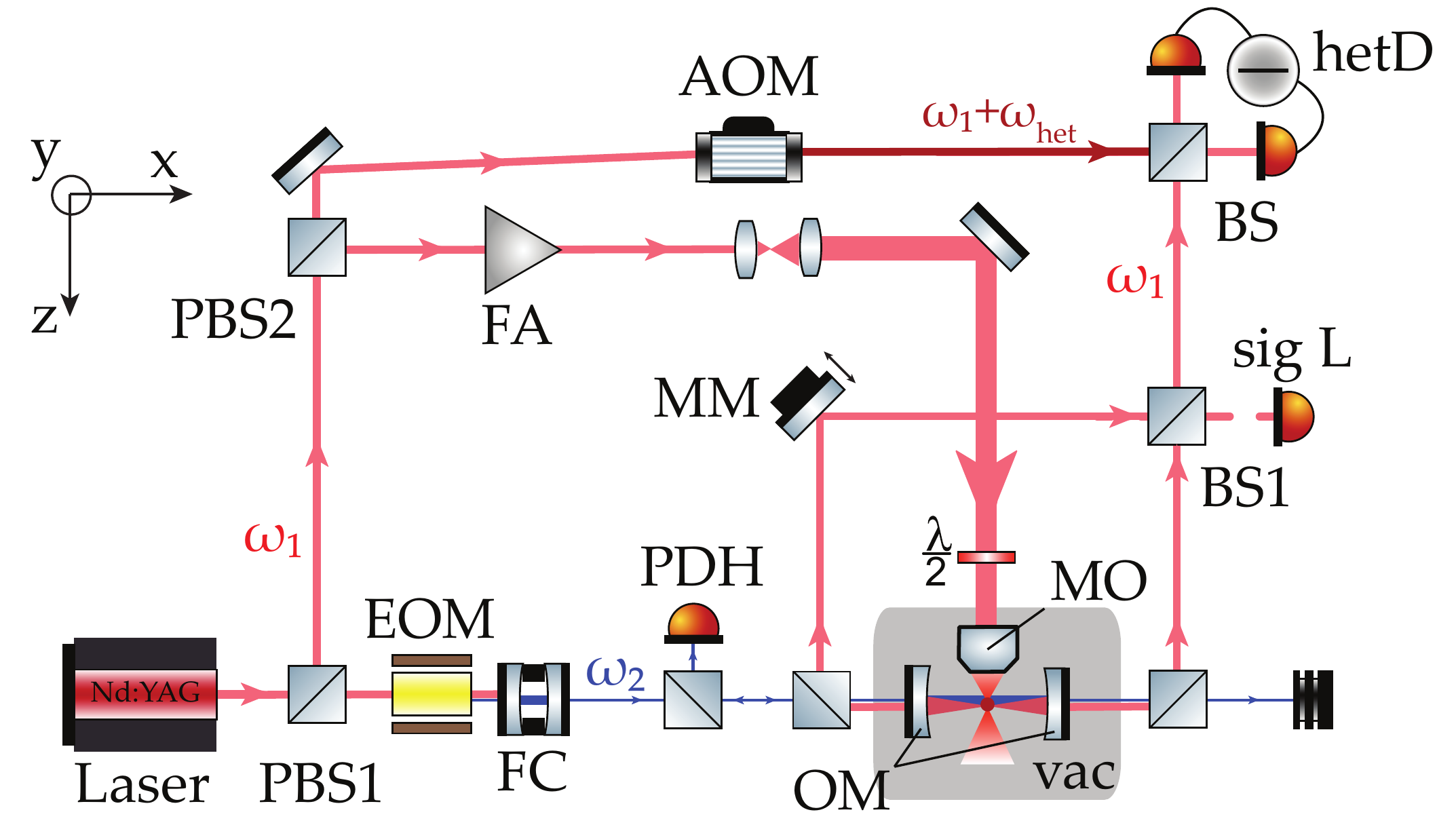}
	\caption{\textbf{The optical setup.} Light from a Nd:YAG master laser (wavelength $\lambda=1064$nm, frequency $\omega_1$) is split into three beams with polarizing beam splitters (PBS1 and PBS2). The first beam (blue) is frequency shifted by a combination of an electro-optical modulator (EOM) and a filtering cavity (FC) to $\omega_2=\omega_1+\Delta\omega_\mathrm{FSR}$, with $\Delta\omega_\mathrm{FSR}$ the free spectral range of the optomechanical cavity (OM). This beam is used for a Pound-Drever-Hall lock (PDH) that stabilizes the master laser on a TEM00 mode of the OM cavity. Note that this scheme automatically stabilizes laser $\omega_1$ to the OM cavity, too. The second beam (red) seeds a fiber amplifier (FA) whose output is used for levitation of a silica nanoparticle in the optical tweezer. We achieve polarization control of the tweezer laser by a half-waveplate ($\lambda/2$) before the microscope objective (MO). The third laser (dark red) is frequency shifted by an acousto-optical modulator (AOM) to $\omega_1+\omega_\mathrm{het}$ and serves as local oscillator for a balanced heterodyne detection (hetD). The microscope objective (MO) sits on a triaxial nano-positioner (not shown) and is used to position the nanoparticle with respect to the OM cavity mode. The OM cavity is driven via photons scattered off of the particle from the tweezer laser. Light leaking out of both cavity mirrors is combined on a beamsplitter (BS1) and used as signal beam for heterodyne read-out (hetD) of the particle motion. To ensure phase stability at BS1 between both beams leaking out of the cavity, a movable mirror (MM) and an error signal derived from one port of BS1 (detector sigL) are used to stabilize their relative phase. 
	}
	\label{fig:setup}
\end{figure}

The optical part of the experiment is based on a Nd:YAG laser (Innolight Mephisto, 2W) with a wavelength of $\lambda=1064$ nm and a frequency $\omega_1=2\pi c/\lambda$ (Fig. \ref{fig:setup}). The laser is split on polarizing beamsplitters (PBS1 and PBS2) into three laser paths, which are used for stabilization to the optomechanical cavity (OM), for optical levitation of a silica nanoparticle and as the local oscillator in heterodyne detection \cite{DelicPhD}.

The first beam (blue) is stabilized to a $\textnormal{TEM}_{00}$ mode of the OM cavity with a Pound-Drever-Hall (PDH) lock. We use an electro-optical modulator (EOM) to generate sidebands at a frequency of $\omega_1\pm \Delta\omega_\mathrm{FSR}$, where $\Delta\omega_\mathrm{FSR}\approx 14.0192$ GHz is the free spectral range of the OM cavity. The filtering cavity (FC) selects only the upper sideband at a frequency $\omega_2=\omega_1+\Delta\omega_\mathrm{FSR}$. The error signal from the PDH detector is used to stabilize the laser frequency $\omega_2$ to the OM cavity resonance $\omega_{cav,2}$ by acting back directly onto the master laser frequency $\omega_1$, thereby automatically stabilizing $\omega_1$ to the adjacent longitudinal cavity resonance frequency $\omega_{cav,1}=\omega_{cav,2}-\Delta\omega_{FSR}$ as well.

The second beam (red) seeds a fiber amplifier (FA, Keopsys, maximal output power 5W) and serves as the trapping laser for the optical tweezer. The optical mode leaving the fiber amplifier is expanded with a telescope, its polarization adjusted with a half-waveplate ($\lambda/2$) and tightly focused with a high NA microscope objective (MO, NA$=0.8$). The microscope objective sits on a triaxial nano-positioner (Mechonics MX35, step size $\sim 8$ nm) that is used to position the levitated nanoparticle within the OM cavity modes. The OM cavity resonance $\omega_{cav,1}$, which is initially empty, is driven from the inside by anti-Stokes and Stokes tweezer photons scattered off the levitated particle at frequencies $\omega_1+\Omega_m$ and $\omega_1-\Omega_m$, respectively ($\Omega_m$ is the mechanical frequency of the particle, where $m=x,y,z$). Note that we can introduce a relative detuning $\Delta$ between the tweezer laser $\omega_1$ and the cavity resonance $\omega_{cav,1}$ by changing the EOM driving tone from $\Delta\omega_\mathrm{FSR}$ to $\Delta\omega_\mathrm{FSR}+\Delta$. The PDH lock will instantaneously stabilize the frequency $\omega_2=\omega_1+\Delta\omega_\mathrm{FSR}+\Delta$ to the cavity resonance $\omega_{cav,2}$, resulting in a detuned tweezer laser with a frequency $\omega_1=\omega_{cav,1}-\Delta$. We use this mechanism to control the optomechanical interaction for cavity cooling.

The third beam (dark red) is frequency shifted by an acousto-optical modulator (AOM) to $\omega_1+\omega_\mathrm{het}$ with $\omega_\mathrm{het}/2\pi=10.2$MHz and used as a local oscillator for a balanced heterodyne read-out of the particle motion (hetD).

Stokes and anti-Stokes scattered photons that leak out of the OM cavity through both mirrors are coherently combined on a beamsplitter (BS1) and used as signal beam for the heterodyne read-out (hetD). A movable mirror (MM) in combination with an error signal (sigL) is used to compensate length fluctuations between the two signal paths (light leaking through the "left" and "right" OM cavity mirror) before superimposition on BS1 \cite{Uberna2010}.

The OM cavity itself has a length of $L\approx1.07$ cm and a linewidth of $\kappa/2\pi\approx193$ kHz. A more detailed description about the interface between OM cavity and tweezer, and positioning of the levitated particle with respect to the OM cavity can be found in \cite{GrassPhD,DelicPhD,Delic2019a}.

\section{Heterodyne detection of the Stokes and anti-Stokes photons}

\begin{figure}[h]
	\includegraphics{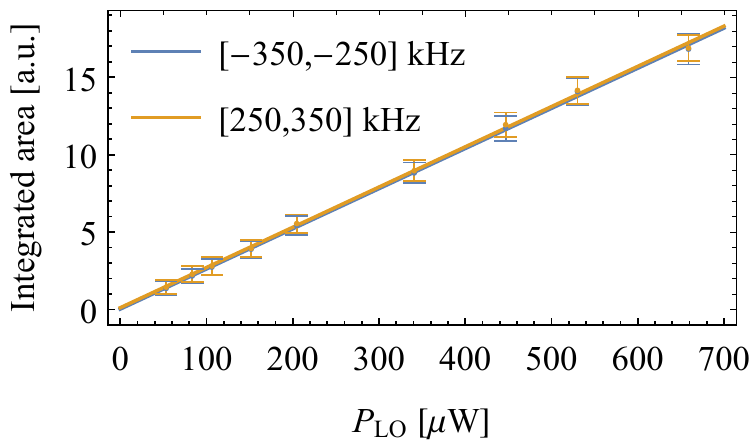}
	\caption{\textbf{Detected sideband power in the relevant frequency band for vacuum signal input.} For different local oscillator powers (LO) and a blocked signal path we integrate the area in a frequency range $(-350, -250)$ kHz (blue) and $(250, 350)$ kHz (orange) around the heterodyne frequency $\omega_\mathrm{het}$, which are relevant to the occupation estimates of the Stokes and anti-Stokes sideband, respectively. The linear scaling of the bandpower with the LO power is a necessary and sufficient signature for shot noise.}
	\label{fig:shotnoise}
\end{figure}

We monitor the particle motion that is coupled to the cavity ($x$ motion) with a heterodyne detection, which allows us to simultaneously detect both Stokes (heating) and anti-Stokes (cooling) photons leaking out of the cavity. These photons are differently amplified by the cavity response, which results in an asymmetry that prefers the anti-Stokes over Stokes photons. Note that cavity cooling of the $x$ motion by coherent scattering comes with a benefit of strong suppression of classical laser intensity and phase noise \cite{Delic2019,DelicPhD}. The local oscillator originates directly from the Mephisto laser such that the added noise by the fiber amplifier is not observed in the fundamental heterodyne spectrum (obtained without the particle signal). We additionally operate the heterodyne at a frequency of $10.2$ MHz and for low local oscillator powers thereby guaranteeing shot-noise limited detection (Fig. \ref{fig:shotnoise}), i.e. the classical intensity noise introduced by the laser is orders of magnitude below the shot noise level \cite{DelicPhD}.

The intrinsic Stokes and anti-Stokes scattered rates $\Gamma_{S}\propto n_x+1$ and $\Gamma_{AS}\propto n_x$ are proportional to the phonon occupation $n_x$. The cavity modifies the scattering rates following the cavity response $T(\Delta,\omega)$ at the particle motional frequency $\Omega_x$:
\begin{equation}
T(\Delta,\omega)=\frac{\left(\frac{\kappa}{2}\right)^2}{\left(\frac{\kappa}{2}\right)^2+(\Delta-\omega)^2},
\end{equation}
where $\kappa$ is the cavity linewidth, $\Delta$ is the detuning of the tweezer laser and $\omega$ is the relevant frequency in a heterodyne spectrum. We observe the modified scattering rates in the heterodyne spectrum $S_\mathrm{het}(\omega)$ with the final ratio of anti-Stokes to Stokes sideband given as \cite{Peterson2016}:
\begin{equation}
\frac{S_\mathrm{het}(\Omega_x)-1}{S_\mathrm{het}(-\Omega_x)-1}=\frac{n_x}{n_x+1}\frac{\left(\frac{\kappa}{2}\right)^2+(\Delta+\Omega_x)^2}{\left(\frac{\kappa}{2}\right)^2+(\Delta-\Omega_x)^2}.\label{hetpeaks}
\end{equation}
Therefore, it is crucial to determine the exact cavity linewidth $\kappa$ and detuning $\Delta$ of the tweezer laser in order to correctly extract the phonon occupation $n_x$, following the procedure described in Ref. \cite{Peterson2016}. After accounting for the cavity response, the remaining asymmetry of the scattering rates is given only by the different probabilities for transitions between adjacent levels and can be used to extract the phonon occupation in the steady state (sideband asymmetry thermometry).

In practice, strongly cooled $x$ motion and weakly cooled $y$ motion typically overlap in the heterodyne spectrum. We evaluate the occupation $n_x$ following three distinct methods with all providing consistent results: from a joint fit of both Stokes and anti-Stokes sidebands of the two peaks, from a fit to only the motional sidebands of the $x$-motion (obtained by generously deleting points around the $y$ peak) and from integrated areas of the data points accredited to the $x$ motion \cite{Underwood2015}. A joint fit of the two motional sidebands, normalized to shot noise, is given by:
\begin{eqnarray}
S_\mathrm{het}(\omega)&=&a_S\frac{\frac{\gamma_x}{2}}{(\omega+\Omega_x)^2+\left(\frac{\gamma_x}{2}\right)^2}\nonumber\\
&&+a_{AS}\frac{\frac{\gamma_x}{2}}{(\omega-\Omega_x)^2+\left(\frac{\gamma_x}{2}\right)^2}+1.\label{fit}
\end{eqnarray}
The ratio of fit amplitudes $a_S$ and $a_{AS}$ is subsequently used to extract the occupation $n_x$ as:
$a_{AS}/a_{S}= n_xT(\Delta,-\Omega_x)/((n_x+1)T(\Delta,\Omega_x))$. 

It has been argued that in the strong cooperativity regime mode hybridization may occur, which would introduce correlations between the $x$- and $y$-motion \cite{Toros2019}. As an ultimate worst case estimate of the phonon occupation, we also integrate the full area of the Stokes and anti-Stokes side in the range $\pm(250-300)$ kHz, thus integrating over spectral peaks of both motions. For the detuning $\Delta/2\pi\approx 315$ kHz we obtain a final phonon occupation of $n_x=0.91\pm 0.04$, which still demonstrates ground state cooling of the particle motion.

\subsection{Cavity linewidth}

\begin{figure}[h!]
	\includegraphics{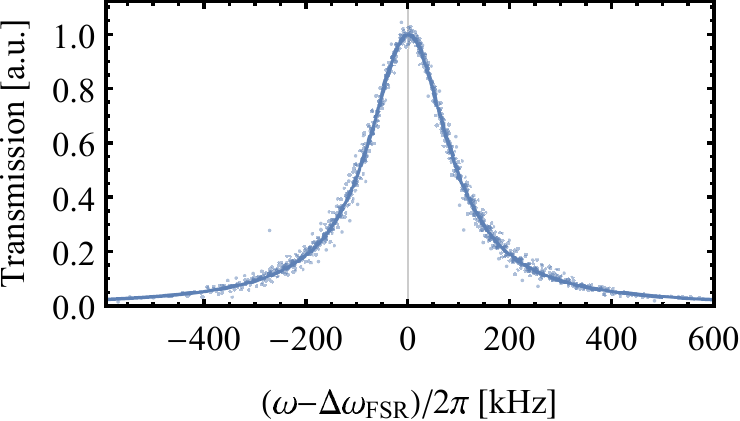}
	\caption{\textbf{Cavity response in transmission.} We regularly scan over a cavity resonance by detuning a laser derived from another cavity mode. Here 26 of these measurements are superimposed to each other with a final cavity decay rate $\kappa/2\pi=(193\pm 4)$ kHz. The average cavity free spectral range from the scans is $\Delta\omega_{FSR}/2\pi=14.0192$ GHz. }
	\label{fig:cavity}
\end{figure}

Here we describe how we determine the cavity linewidth $\kappa$. The principal locking scheme is described in the text above. Instead of driving the cavity with the tweezer scattered photons, we now drive the cavity resonance $\omega_{cav,1}$ with the laser at a variable frequency $\omega_1=\omega_{cav,1}-\Delta$ through one of the OM cavity mirrors. We tune the detuning $\Delta$ over the cavity resonance and monitor the laser power in transmission of the OM cavity (Fig. \ref{fig:cavity}):
\begin{equation}
P_{tran}\propto T(\Delta)=\frac{\left(\frac{\kappa}{2}\right)^2}{\left(\frac{\kappa}{2}\right)^2+\Delta^2}.
\end{equation}
As this procedure is regularly repeated, we combine all measurements to get an average cavity linewidth $\kappa/2\pi=(193\pm4)$ kHz. We extract the mean free spectral range $\Delta\omega_{FSR}/2\pi\approx 14.0192$ MHz from the scan as well, which allows us to calculate the cavity length as $L=c\pi/\Delta\omega_{FSR}\approx 1.07$ cm.

\subsection{Detuning of the tweezer laser}

We use the cavity induced asymmetry of the residual phase noise and the spectrum of weakly cooled $z$-motion in the heterodyne detection to extract the detuning of the tweezer laser. The ratio of the spectra on each side of the heterodyne frequency, as its source is purely classical, is given by:
\begin{equation}
\frac{S_\mathrm{het}(\omega)}{S_\mathrm{het}(-\omega)}=\frac{\left(\frac{\kappa}{2}\right)^2+(\omega+\Delta)^2}{\left(\frac{\kappa}{2}\right)^2+(\omega-\Delta)^2},
\end{equation}
where $\Delta$ is the tweezer laser detuning. The typical error of this method is at maximum $\pm 10$ kHz, which is used in the error estimation of the phonon occupation.

\section{Positioning stability}

\begin{figure}[h]
	\includegraphics{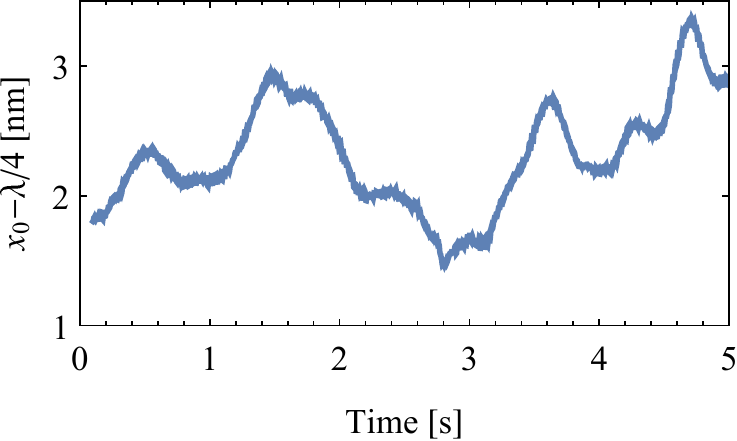}
	\caption{\textbf{Particle distance from the cavity node.} We position the nanoparticle in the vicinity of the cavity node. The absolute position is calibrated from the intracavity photon number for a particle placed at the cavity antinode. The particle is at most $\sim 4$ nm away from the cavity node, which strongly suppresses elastic scattering into the cavity and phase noise, leading to low phase noise heating.}
	\label{fig:position}
\end{figure}

The peak size in the heterodyne measurement at the heterodyne frequency $\omega_\mathrm{het}$ is given by the product of the local oscillator power $P_{LO}$ and the optical power leaking out of the cavity $P_{cav}\propto n_{phot}$, where $n_{phot}$ is the intracavity photon number. It depends on the nanoparticle position $x_0$ along the cavity standing wave as $n_{phot}\propto \cos^2(kx_0)$. Therefore, the height of the heterodyne peak holds information about the positioning stability of the nanoparticle. In order to calibrate the absolute particle position, we initially place the nanoparticle around the cavity antinode ($x_0=0$) and monitor the peak height, with the maximum peak size subsequently rescaled to $n_{phot}=1$. The minimum intracavity photon number $n_{phot}=0$ is reached at a cavity node ($x_0=\lambda/4=256$ nm), where the optimal cooling of the $x$ motion takes place. We assume no changes in the optical power of the local oscillator and the detection efficiency during this measurement. Using this calibration we are able to follow the particle position in the vicinity of the cavity node during cavity cooling, which shows that the particle stays within a couple of nanometers (Fig. \ref{fig:position}). Besides long term drifts, the most of the particle motion around the cavity node comes from the vibrations transferred from the turbo pump, which are at a frequency of $\omega_{jitter}/2\pi=54$ Hz (Note that in Figure \ref{fig:position} this jitter is not seen as we average over it to show only long term positioning stability). This is far away from any relevant mechanical frequencies and doesn't induce additional heating of the particle motion.

\section{Noise sources}

\subsection{Laser phase noise}

By using the cavity drive deduced from the cooling performance in Ref. \cite{Delic2019} and accounting for the increased tweezer power to reach higher mechanical frequencies, we calculate the cavity drive of $E_d/2\pi\approx 4\times10^9$ Hz. The resulting intracavity photon number is subsequently given by:
\begin{equation}
n_{phot} (x_0)=  \frac{E_d^2\cos^2(kx_0)}{\left(\frac{\kappa}{2}\right)^2+\Delta^2},
\end{equation}
where $x_0$ is the particle position along the cavity standing wave.
The expected number of phonons added from classical laser phase noise at the optimal detuning $\Delta\sim \Omega_x$ and for a particle positioned $3$ nm away from the cavity node at $\lambda/4$ is \cite{Rabl2009,Kippenberg2013,Safavi-Naeini2013}:
\begin{equation}
n_{phase}=\frac{n_{phot} (\lambda/4+3\textnormal{nm})}{\kappa}S_{\dot{\phi}\dot{\phi}} (\Omega_x )< 0.025,
\end{equation}
where the measured laser frequency noise of a Mephisto laser at the motional frequency $\Omega_x/2\pi=305$ kHz is $S_{\dot{\phi}\dot{\phi}} (\Omega_x)< 0.1 \textnormal{Hz}^2/\textnormal{Hz}$ \cite{JasonPhD}. This corresponds to a phase-quadrature noise contribution of maximally $C_{pp}=2.5\times 10^{-3}$ in excess to fundamental vacuum fluctuations \cite{Sudhir2017}. 

\subsection{Laser intensity noise}

\begin{figure}[h!]
	\includegraphics[width=\columnwidth]{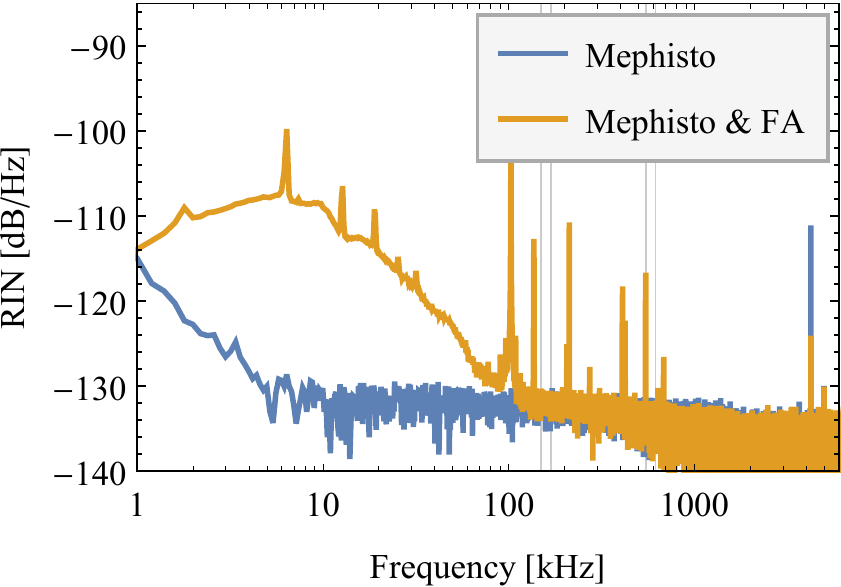}
	\caption{\textbf{Laser intensity noise of the tweezer laser.} Laser intensity noise at twice the mechanical frequency leads to parametric heating. We measure the relative intensity noise (RIN) from the Mephisto laser (blue) and the combined noise with the fiber amplifier used for trapping (orange). The added intensity noise of the fiber amplifier is due to the limitations of the amplifier's current locking circuit. However, at the relevant frequency band (shaded gray area) the intensity noise is at the level of the noise intrinsic to the Mephisto laser.}
	\label{fig:intnoise}
\end{figure}

Intensity noise of the tweezer laser leads to parametric heating due to the modulation of the trap intensity and manifests itself as a negative damping rate \cite{Savard1997}:
\begin{equation}
\gamma_{int}^m=-\pi^2 \Omega_m^2 S_{RIN}(2\Omega_m).
\end{equation}
Here, $S_{RIN}(2\Omega_m)$ is the relative intensity noise (RIN) at twice the mechanical frequency $\Omega_m$. While the intensity noise of a Mephisto laser is typically around $-135~ \textnormal{dB}/\textnormal{Hz}$, the fiber amplifier used to increase the trapping power significantly increases the laser intensity noise at frequencies below $100$ kHz (Fig. \ref{fig:intnoise}). We therefore intentionally increase the trapping power such that second harmonics of the motional frequencies are in the range of low laser intensity noise. The calculated damping rates are $(\gamma_{int}^x,\gamma_{int}^y,\gamma_{int}^z)/2\pi=-(6,5,0.5)$ mHz. Heating is therefore negligible for the strongly cooled $x$ and $y$ motion with cooling rates of at least $1$ kHz (see below). However, for the only sporadically cooled $z$ motion, the heating rate can become comparable to the cooling rate. 

Laser intensity noise can also add to the phonon occupation through cavity backaction \cite{Sudhir2017,Jayich2012}. For our parameters at optimal detuning, the corresponding amplitude-quadrature noise contribution yields $C_{qq}=2\times 10^{-5}$ in units of tweezer laser shot noise. Therefore, for a particle placed $4$ nm away from the cavity node this adds at maximum $n_{int}\sim 10^{-4}$ phonons to the occupation $n_x$ \cite{Sudhir2017}. As in the case of phase noise, the strong suppression of intracavity photons renders the noise contribution negligible.

\section{Heating rates}

\begin{figure}[h!]
	\includegraphics[width=0.75\columnwidth]{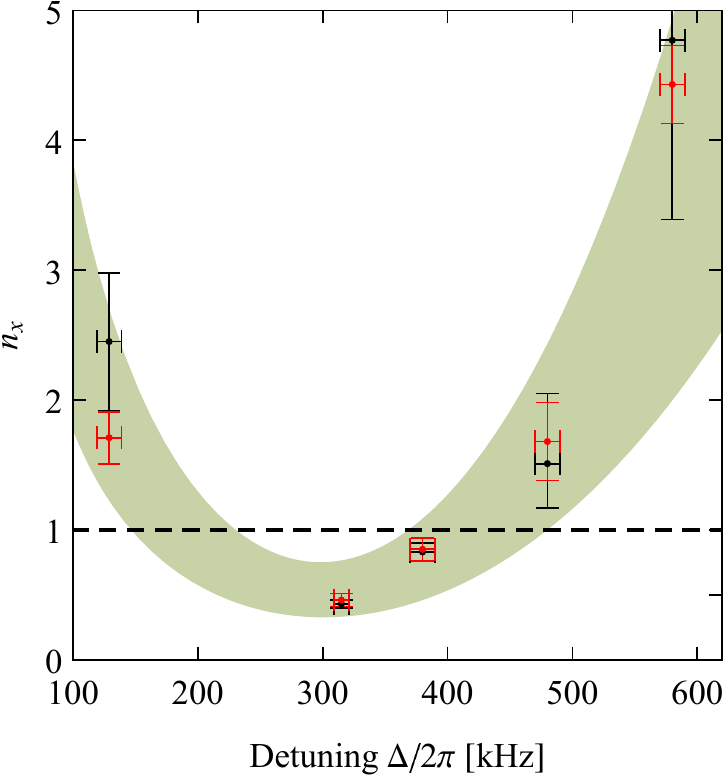}
	\caption{\textbf{Phonon occupation deduced from sideband thermometry (black) and from heating and cooling rates (red).} We measure the phonon occupation from sideband asymmetry, which does not depend on detection calibration, and compare it to the occupation obtained from the independently measured heating and cooling rates. The occupations obtained from these two methods are in a good agreement.}
	\label{fig:crosscheck}
\end{figure}

In our system, the heating rate is dominated by gas collisions at the relevant pressures of $\sim 10^{-6}$ mbar. This was studied in detail in Ref. \cite{Delic2019a} in the same experimental apparatus and using particles of the same nominal diameter. This allows us to extrapolate the gas heating rate at any pressure and for any mechanical frequency (or trapping power). For our experimental parameters at the detuning of $\Delta/2\pi\approx 315$ kHz, the gas heating rate is $\Gamma_{gas}/2\pi=(16.1\pm 1.2)$ kHz, which we subtract from the inferred heating rate $\Gamma_x/2\pi=(20.6\pm 2.3)$ kHz to estimate the recoil heating rate $\Gamma_{rec}/2\pi=4.3\pm 1$ kHz. This is in a good agreement with our calculated recoil heating $\Gamma_{rec,calc}/2\pi=6$ kHz, which we obtain from a model of the trapping potential based on experimental mechanical frequencies \cite{DelicPhD, NovotnyBook} (note that a direct measurement of the recoil heating rate would require operating at much lower pressures \cite{Jain2016}). Together with the  independently measured cooling rates $\gamma_x$, which are obtained from the fits to the mechanical sidebands using Eq. \ref{fit}, we can estimate the phonon occupation via $n_x = (\Gamma_{gas} + \Gamma_{rec,calc})/\gamma_x$. The thus obtained values are in good agreement with the measurements obtained by sideband asymmetry (Figure \ref{fig:crosscheck}). We use the full optomechanical theory from Ref. \cite{Genes2008} to describe the cooling performance of our system and plot the theoretical green band shown in Figure 2 of the main text. In addition to independently measured system parameters, the upper and lower bound of the band is given by the range of experimental pressures during our measurements.

\section{Cavity cooling of other motional axes}

In principle, cavity cooling by coherent scattering enables cooling in all three dimensions \cite{Delic2019}. In our current experiment, however, we optimize cooling only along the cavity ($x$-)axis, specifically by aligning the tweezer polarization orthogonal to the cavity axis (thereby maximizing the scattering component along the $x$-axis), and by positioning the particle at the cavity node. We explore here the cooling of the $y$ and $z$ motion.

\subsection{Cavity cooling of the z motion}

There are two distinct mechanisms by which the $z$ motion is coupled to the cavity mode:
\begin{enumerate}
	\item Cavity cooling by coherent scattering is a genuine three-dimensional cooling process. The optimal position for the cooling of the $z$ motion is at the cavity antinode \cite{Delic2019}, while at the cavity node the $z$ motion is not cooled at all. However, the particle is always positioned in the vicinity of the cavity node and thus experiences some minimal cooling of the $z$ motion.
	\item The tweezer laser is slightly non-orthogonal to the cavity mode, which leads to a projection of the $z$ motion onto the cavity axis. This results in some cooling of the $z$ motion even at the cavity node.
\end{enumerate}

Harmonics of the $z$ motion in the heterodyne detection are present due to the nature of coupling to the cavity mode \cite{DelicPhD}. For a particle at any position $x_0$, the area of the $i$-th harmonic in the heterodyne spectrum is proportional to the $2i$-th order central moment $\langle z^{2i}\rangle$ of the $z$ motion. The ratio of the first and second harmonic after removing the cavity response from the heterodyne detection is:
\begin{equation}
\frac{g_{quad}}{g_{lin}}\frac{\langle z^{4}\rangle}{\langle z^{2}\rangle}=\frac{k_B T_z}{m\Omega_z^2}\left(\frac{k-1/z_R}{2}\right)^2,
\end{equation}
where $m=2.83$ fg is the nanoparticle mass \cite{Delic2019a}, $k=2\pi/\lambda$ with a wavelength $\lambda=1064$ nm, $z_R=W_xW_y\pi/\lambda$ is the Rayleigh length with trap waists $W_x=0.67~\mu\text{m}$ and $W_y=0.77~\mu\text{m}$ and $T_z$ is the temperature of the $z$ motion. For the tweezer detuning $\Delta/2\pi=315$ kHz the average observed temperature is $T_z\approx 80$ K and the respective occupation is $n_z=\frac{k_BT_z}{\hbar\Omega_z}=2\times 10^7$. The temperature greatly varies between different measurements, but the particle is cooled enough along the $z$-axis to be stably trapped in high vacuum.

\subsection{Cavity cooling of the y motion}

The optical potential created by the tweezer is, due to the tight focusing, elliptical in the focal plane \cite{NovotnyBook}. As a result, the particle mechanical frequencies along the $x$- and $y$-axes are nondegenerate. The orientation of the elliptical trap in the focal plane can be aligned by rotating the polarization of the tweezer laser. In this experiment, we optimize the polarization such that the $y$ motion is orthogonal to the cavity axis. However, due to experimental imperfections, there remains always a small projection of the $y$ motion onto the cavity axis. Since the particle position is optimized for maximal coupling to the motion along the cavity axis, the $y$ motion is also coupled to the cavity axis and hence is also moderately cooled by coherent scattering. We estimate the occupation number by computing the power spectral density of the heterodyne detection and fitting the motional peak of the $y$ motion to extract the damping rate of at least $\gamma_y/2\pi=0.5$ kHz. Together with the calculated heating rate $\Gamma_y/2\pi \approx 20$ kHz, dominated by gas collisions and recoil heating of the trapping laser, we can estimate the occupation of the $y$ motion to be well below $n_y < 100$ phonons for any detuning.

\section{Decoherence rate}

We estimate the decoherence in our experiment after preparing the nanoparticle in its ground state of motion and during a subsequent hypothetical free fall based on Romero-Isart et al.\cite{ORI2011}.

The heating rate of $\Gamma_x/2 \pi=20.6 \pm 2.3$~kHz is given by the optical recoil heating rate $\Gamma_{rec}/2 \pi\approx 6$~kHz and recoil by background gas collisions $\Gamma_{gas}/2 \pi\approx 15$~kHz. It limits the coherence time in the optical trap to $T_{\mathrm{trap}}=1/\Gamma_x=7.6 \pm 1~\mu$s corresponding to approximately 15 coherent oscillations before the ground state is populated with one phonon.

When the optical trap is turned off, the particle is in free fall and coherently expands. The major decoherence mechanism is then position localization via background gas collisions. We are operating in an interesting transitional regime, in which the initial wavepacket size $x_{\mathrm{zpf}}=\sigma(t=0)=\sqrt{\frac{\hbar}{2 m \Omega_x}}=3.1~$pm is smaller than the thermal de Broglie wavelength of the scattering gas molecules $\lambda_{th}=2a=h/\sqrt{m_\mathrm{gas} k_B T}=19$~pm by a factor of $~6.2$ ($T$: environment temperature, $m_\mathrm{gas}\approx 28u$: mass of one gas molecule, $u=1.67\times10^{-27}$ kg: atomic mass unit).

In the short-distance regime where $\sigma(0) \ll 2a$ the decoherence rate increases quadratically with the wavepacket size (and hence with time due to the linear expansion) until it saturates in the regime $\sigma(t) \gg 2a$. To estimate the expected coherence time and maximum expansion of the wavepacket, we can determine the localization parameter from our experimental heating rate in the optical trap $\Lambda=\Gamma_x/x_{\mathrm{zpf}}^2$. This is obtained by comparing the energy increase by momentum diffusion (Eq. 14 in \cite{ORI2011}) with the heating rate: $\frac{2 \Lambda \hbar^2}{2 m}=\Gamma_x \hbar \Omega_x$. From this, we can determine the maximum coherence length $\xi_{max}=10.2~$pm and the corresponding expansion time $t_{max}=1.42~\mu$s in the short-distance approximation (Eq. 19 and 20 in \cite{ORI2011}):
\begin{eqnarray}
t_{max}&=&\left(\frac{3m\left(2 \bar{n}_x +1\right)}{2\Lambda \hbar \Omega_x} \right)^{1/3}\nonumber\\
\xi_{max}&=&\sqrt{2}\left(\frac{2 \hbar \Omega_x}{3 m \Lambda^2 (2 \bar{n}_x +1)} \right)^{\frac{1}{6}}
\end{eqnarray}
where $\bar{n}_x\approx 0.43$ is the inital occupation of the wavepacket. Note that the obtained values slightly underestimate the expected values, as the expansion occurs in the transition to saturation.

What conditions are required to achieve an expansion of the wavepacket until it reaches the size of the nanosphere itself? Most of this evolution occurs with a wavepacket size $\sigma(t)$ that is much larger than the deBroglie wavelength of the gas molecules and thus in the long-distance approximation. Here, the decoherence rate is given by $\Gamma_{sat}=\lambda_{th}^2 \Lambda=\Gamma_x \lambda_{th}^2/x_\mathrm{zpf}^2 \approx 3.6~$MHz (note that $\Gamma_{sat}$ is denoted $\gamma$ in \cite{ORI2011}) and essentially corresponds to the rate of collision with single gas molecules, as in the long-distance approximation already a single collision localizes the nanoparticle.

Given the expansion of the undisturbed wavepacket via $\sigma(t)=x_\mathrm{zpf}(1+\Omega_x t)$ we require an expansion time of $\tau \approx r/(x_\mathrm{zpf} \Omega_x)=12$~ms, demanding a decoherence rate below 84~Hz. This is achievable by a reduction of the pressure by at least a factor of $5 \times 10^4$ to below $2 \times 10^{-11}$~mbar. However, at these pressures blackbody radiation of the internally hot particle becomes relevant. The total localization parameter due to blackbody radiation consists of three contributions: absorption $\Lambda_{bb,a}$, emission $\Lambda_{bb,e}$ and scattering $\Lambda_{bb,sc}$ \cite{ORI2011}:
\begin{eqnarray}
\Lambda_{bb}&=&\Lambda_{bb,sc}+\Lambda_{bb,e}+\Lambda_{bb,a}\approx 2.3\times 10^{20}~\textnormal{Hz}/\textnormal{m}^2, \textnormal{with:}\nonumber\\
&&\Lambda_{bb,sc}\approx 10^{15}~\textnormal{Hz}/\textnormal{m}^2\nonumber\\
&&\Lambda_{bb,e(a)}\approx  2.3\times 10^{20}~\textnormal{Hz}/\textnormal{m}^2(1.4\times 10^{18}~\textnormal{Hz}/\textnormal{m}^2).\nonumber\\
\end{eqnarray}
Here we assumed an internal temperature of $700$ K, which is a conservative upper bound based on measured heating rates in Ref. \cite{Delic2019a}, an environment temperature of $300$ K and an average refractive index of $\epsilon_{bb}=2.1+i\times0.57$. The expansion time $t_{max,bb}\approx 0.55~$ms and length $\xi_{max,bb}\approx 2~$nm is therefore limited by the blackbody radiation. To further reduce decoherence to the desired level (below the gas scattering contribution) requires cryogenic temperatures (below 130K) for both the internal particle temperature and the environment. This could be achieved either by combining a cryogenic (ultra-high) vacuum environment with laser refrigration of the nanoparticle \cite{Rahman2017} or with low-absorption materials \cite{Frangeskou2018}.

\makeatletter
\renewcommand*{\@biblabel}[1]{\hfill#1.}
\makeatother

\end{document}